\documentclass[prl,final,twocolumn,showpacs]{revtex4}
\usepackage{graphicx}

\begin{document}

\title[The Andreev states of a superconducting quantum dot]{The Andreev states of a superconducting quantum dot: mean field vs exact numerical results}

\author{A. Mart\'{\i}n-Rodero and A. Levy Yeyati}
\address{Departamento de F\'{i}sica Te\'orica de la Materia
Condensada and Instituto Nicol\'as Cabrera, 
Universidad Aut\'{o}noma de Madrid, E-28049 Madrid, Spain}

\begin{abstract}
We analyze the spectral density of a single level quantum dot coupled to
superconducting leads focusing on the Andreev states appearing within
the superconducting gap. We use two complementary approaches: the numerical
renormalization group and the Hartree-Fock approximation. Our results 
show the existence of up to four bound states within the gap when the
ground state is a spin doublet ($\pi$ phase). Furthermore the results 
demonstrate the reliability of the mean field description within this phase.
This is understood from a complete correspondence that can be established
between the exact and the mean field quasiparticle excitation spectrum 
within the gap.
\end{abstract}
\maketitle

\section{Introduction}

Much progress has been achieved in recent years on the transport properties of quantum dots
coupled to superconducting leads (for a review see \cite{us2011}). A central concept in the understanding of
these properties is that of Andreev bound states (ABS), i.e. the bound states appearing within the 
superconducting gap due to multiple Andreev reflections at the dot superconductor interfaces.
When the two leads are superconducting the ABS depend on the superconducting phase difference
and are thus current carrying states (typically most of the Josephson current is carried by the ABS \cite{us2011}).
The interest in the ABS in these kind of systems has been increased by recent 
experiments allowing
for their direct measurement through tunnel spectroscopy on carbon nanotube and graphene quantum dots \cite{pillet2010,dirks}. 
The issue is also related to the strong activity in the search of Majorana fermions, which would manifest as
midgap states in different type of hybrid nanostructures \cite{alicea}.

The simplest situation for analyzing the ABS spectrum is that of a single quantum dot (QD)
with large energy level spacing which can be appropriately described by the single level Anderson model \cite{anderson61}. The presence of subgap states for the
case of a magnetic impurity in a BCS superconducting host was already 
demonstrated in Refs. \cite{shiba,jarrel}.
This model was afterward extended to analyze the Josephson transport properties
and the so called 0-$\pi$ transition which signals the transition from a 
singlet ($S=0$) to a doublet ($S=1/2$) ground state 
\cite{glazman,arovas,vecino,choi2,siano,oguri,ansari}. 
In spite of these theoretical efforts 
the knowledge about the detailed structure of the ABS spectrum appears
to be somewhat disperse in the literature.
Thus, in a method like the non-crossing approximation (NCA), which
is able to describe the 0-$\pi$ transition in the large charging energy
regime, a single subgap resonance appears whose
crossing of the Fermi level signals the transition \cite{sellier}. 
On the other hand,
other approximations like  Hartree-Fock (HFA) or exact diagonalizations 
in the infinite
$\Delta$ limit, where $\Delta$ denotes the superconducting gap parameter 
in the leads, point to the existence of up to 4 levels symmetrically located around the Fermi energy in the $\pi$-phase \cite{vecino,simon,luitz}. 
The bound state spectrum has also been analyzed using
the numerical renormalization group
(NRG) method \cite{choi,hewson,hecht} 
although in these works only up to two ABS 
were identified. 
A previous NRG calculation showing up to four ABS exists
\cite{japs} although in this work only the single lead case was considered
(i.e. without a phase difference between the leads).
Taking into account all this rather fragmented 
evidence it seems worthwhile to investigate this issue in more detail using 
numerically exact results compared to different approximations. This is further
motivated by the possibility of a direct experimental test along the lines
of recent works \cite{pillet2010,dirks}.

In the present work we give a detailed analysis of the ABS for the 
single level Anderson model
coupled to superconducting leads focusing in the regime $T_K < \Delta$,
where $T_K$ denotes the Kondo temperature, which appears to be the relevant
one for describing the experimental results of Ref. \cite{pillet2010}.
We use NRG calculations and compare the results with the mean-field
approach provided by the HFA. It is found that when the system undergoes the 
transition to the 
$\pi$-phase there appear in general up to 4 ABS in agreement with the analysis of the simple
spin-polarized HFA \cite{vecino}. Indeed, our analysis shows that the HFA provides a quite fair
description of the bound state spectra except in the regime where Kondo correlations dominate over
the superconducting ones.

The rest of the paper is organized as follows: in Sect. \ref{model} we describe the model used
for a QD coupled to superconducting leads and the basic theoretical analysis of its spectral properties;
in Sect. \ref{S-QD} we consider the simpler situation of a QD coupled to a single superconducting electrode
(S-QD case) and present results for the ABS spectrum using both NRG and HFA, analyzing the range of
validity of this last approximation for this case. In Sect. \ref{S-QD-S} this analysis is extended to 
a phase-biased S-QD-S system where we study in particular the behavior of the ABS spectrum around
the transition between singlet and doublet ground states. Finally in Sect. \ref{conclusions} we
give some concluding remarks.

\section{Model and basic theoretical analysis}
\label{model}

A minimal model for a QD coupled to metallic electrodes in the regime
where the energy level spacing $\delta \epsilon$ is sufficiently large to restrict the analysis to
a single spin-degenerate level is provided by the single level Anderson model \cite{anderson61},
with the Hamiltonian $H=H_L+H_R+H_T+H_{QD}$ where $H_{QD}$ corresponds
to the uncoupled dot given by
\begin{equation}
H_{QD} = \sum_{\sigma} \epsilon_0 c^{\dagger}_{0\sigma} c_{0\sigma} +
U n_{0\uparrow} n_{0\downarrow} ,
\label{HQD}
\end{equation}
where $c^{\dagger}_{0\sigma}$ creates and electron with spin $\sigma$ on the dot level
located at $\epsilon_0$ and $U$ is the local Coulomb interaction for two electrons
with opposite spin within the dot ($n_{0\sigma}= c^{\dagger}_{0\sigma}c_{0\sigma}$).
On the other hand, $H_{L,R}$ describe the uncoupled left and right leads which are
superconductors represented by a BCS Hamiltonian of the type
\begin{equation}
H_{\nu} = \sum_{k\sigma} \xi_{k,\nu} c^{\dagger}_{k\sigma,\nu} c_{k\sigma,\nu} +
\sum_{k} \left(\Delta_{\nu} c^{\dagger}_{k\uparrow,\nu}c^{\dagger}_{-k\downarrow,\nu} +
\mbox{h.c.} \right),
\label{Hnu}
\end{equation}
where $c^{\dagger}_{k\sigma,\nu}$ creates an electron with spin $\sigma$ at the
single-particle energy level $\xi_{k,\nu}$ of the lead $\nu=L,R$
(usually referred to the lead chemical potential, i.e. $\xi_{k,\nu} = \epsilon_{k,\nu} - \mu_{\nu}$)
and $\Delta_{\nu}=|\Delta_{\nu}|\exp{(i\phi_{\nu})}$ is the (complex) superconducting order parameter on lead $\nu$.
Finally, $H_T$ describes the coupling between the QD level to the leads and has the form
\begin{equation}
H_{T} = \sum_{k\sigma,\nu} \left(V_{k,\nu} c^{\dagger}_{k\sigma,\nu} c_{0\sigma} + \mbox{h.c.} \right).
\label{HT}
\end{equation}

The coupling to the leads is usually characterized by a single parameter 
$\Gamma_{\nu} = \pi \rho_{\nu} |V_{\nu}|^2$, determining the width of the one-electron resonance. In this expression $V_{\nu}$ corresponds to an average
over the Fermi surface of $V_{k,\nu}$ and $\rho_{\nu}$ denotes the
corresponding density of states on the leads. 

In this work we are interested in the spectral properties which can be extracted from the
dot Green's functions defined as $\hat{G}^{r}_{\sigma}(t,t') = -i\theta(t-t') \langle \left[\Psi_{\sigma}(t)
,\Psi^{\dagger}_{\sigma}(t')\right]_+\rangle$, where 
$\Psi_{\sigma} = ( c_{0\sigma} \; c^{\dagger}_{0,-\sigma})^T$ and 
the average is taken over the system ground state. 
In the case of the Anderson model with superconducting leads two
types of ground states appear depending on the parameters: a non-degenerate
$S=0$ ground state (0 phase) or a double-degenerate $S=1/2$ ground state
($\pi$-phase). The Green's functions can be formally written
in frequency space using the Lehmann representation

\begin{widetext}
\begin{equation} 
\hat{G}^{r}_{\sigma,S_z}(\omega) =  \sum_m
\frac{\langle \Phi_{0,S_z} | \Psi_{\sigma} |\Phi_m \rangle
\langle \Phi_m | \Psi^{\dagger}_{\sigma} |\Phi_{0,S_z} \rangle}{\omega 
- \left(E_m - E_0 \right) + i0^+} + 
\frac{\langle \Phi_{0,S_z} | \Psi^{\dagger,T}_{\sigma} |\Phi_m \rangle
\langle \Phi_m | \Psi_{\sigma}^{T} |\Phi_{0,S_z} \rangle}{\omega 
+ \left(E_m - E_0 \right) + i0^+} ,
\label{lehmann}
\end{equation}
\end{widetext}

where the system ground state, denoted by $|\Phi_{0,S_z}\rangle$, 
maybe degenerate ($S_z=\pm1/2$) and
$m$ labels the excited states having an extra quasiparticle with
respect to the ground state. In the degenerate case the total Green
function is finally obtained as
$\hat{G}^r_{\sigma} = \frac{1}{2} \sum_{S_z} \hat{G}^r_{\sigma,S_z}.$
The formal expression of Eq. (\ref{lehmann}) allows a direct calculation of the quasiparticle
spectral densities using numerical methods like the NRG method which we
describe further below. 

It would be interesting to compare the NRG results for the spectral 
density with those provided
by different approximations specially with HFA
which could provide a rather simple scheme for describing recent
experimental results on the ABS spectrum \cite{pillet2010}.
In this approximation the dot Green's function is given by
$(\hat{G}^{r,HF}_{\sigma})^{-1} = (\hat{G}^{r,(0)})^{-1} - 
\hat{\Sigma}^{HF}_{\sigma}$,
where $\hat{G}^{r,(0)}$ is the non-interacting dot Green's function in Nambu
space and the self-energy $\hat{\Sigma}^{HF}_{\sigma}$ corresponds to the
first order diagrams in the Coulomb interaction and are given by

\begin{eqnarray} 
\Sigma^{HF}_{11,\sigma} &=& -\Sigma^{HF}_{22,-\sigma}= U <n_{0,-\sigma}> 
\nonumber\\
\Sigma^{HF}_{12,\sigma} &=& (\Sigma^{HF}_{21,\sigma})^* = -U <c_{0\uparrow} c_{0\downarrow}>
\end{eqnarray}

In the HFA both $<n_{0,\sigma}>$ and $<c_{0\uparrow} c_{0\downarrow}>$ have to
be calculated self-consistently. The explicit expression for $\hat{G}^{r,HF}$ is 
\begin{widetext}
\begin{equation}
\hat{G}^{r,HF}_{\sigma} = \left( \begin{array}{cc} \omega - \epsilon_0 -
\Gamma g(\omega) - \Sigma^{HF}_{11,\sigma} & 
\Gamma \cos{\frac{\phi}{2}} f(\omega) - \Sigma^{HF}_{12,\sigma} \\
\Gamma \cos{\frac{\phi}{2}} f(\omega) - \Sigma^{HF}_{21,\sigma} & 
\omega + \epsilon_0 - \Gamma g(\omega) - \Sigma^{HF}_{22,\sigma}
\end{array} \right)^{-1} 
\label{hfa}
\end{equation}
\end{widetext}
where $\Gamma_L = \Gamma_R = \Gamma/2$, $\phi$ denotes the superconducting-phase difference 
and $g(\omega) = -\Delta f(\omega)/\omega = 
-\omega/\sqrt{\Delta^2 - \omega^2}$ 
are the dimensionless BCS Green's functions of
the uncoupled leads. 

\begin{figure}[!t]
\begin{center}
\includegraphics[scale=.4,angle=0]{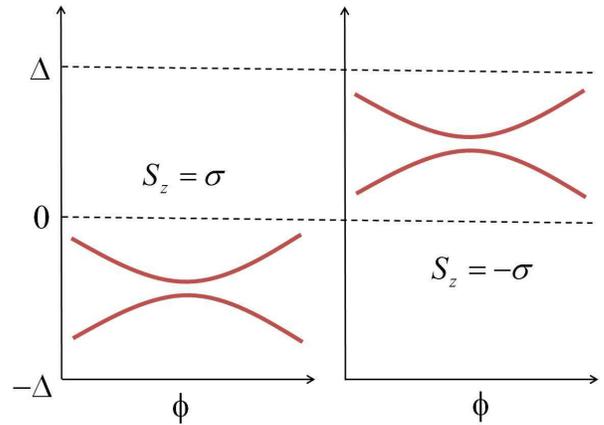}
\caption{(Color online) Subgap pole structure of the Green's functions
components $G_{\sigma,S_z}(\omega)$ in the $\pi$-phase for the
two different orientations of the ground state spin $S_z$.
As discussed in the text these excitations have either a purely
hole-like or electron-like character, which allows a correspondence
with the results of the HFA. Spin-symmetry is recovered when
taking the trace over the degenerate ground state.} 
\end{center}
\label{scheme-ABS}
\end{figure}

Within the HFA the transition from the 0 to the $\pi$ phase 
\cite{shiba,arovas,vecino} is signaled by the existence of a spin 
broken symmetry 
solution with $<n_{0,\sigma}> \neq <n_{0,-\sigma}>$. Although this symmetry breaking
is not actually present in the exact solution,
the HFA provides a very accurate description of the ABS spectrum within 
the $\pi$ phase as will be shown along this work. 
This can be qualitatively understood by analyzing the low energy 
pole structure of the exact Green's function given by Eq. (\ref{lehmann}) 
when the system is in the $\pi$-phase. In this case the two-fold degenerate 
ground state can be labeled by $S_z=\pm1/2$. Quasiparticle excitations over
this ground state can correspond to transitions to states with total spin
either $S=0$ or $S=1$. However, in the last case the excitation energy is
necessarily larger than $\Delta$ as these excited states involve an unpaired
electron in the leads. Therefore the excitations within the gap can only 
arise from transitions to states with total spin equal to zero.
It is then straightforward to see 
that subgap electron-like excitations with spin 
$\sigma$ can only be created from the ground state with $S_z=-\sigma$ while the 
hole-like excitations arise from the ground state $S_z=\sigma$. 
This structure is illustrated in Fig. \ref{scheme-ABS} where we show 
schematically the subgap poles in $G_{\sigma,S_z}(\omega)$ 
in the $\pi$-phase for $S_z = \pm 1/2$. 
This is precisely the structure of the
subgap excitations which are found in the spin-polarized HFA: the solutions
for the majority and minority spin populations have 
only hole-like or electron-like character respectively.
Therefore one can establish a correspondence between the
exact and the HFA excitations for the subgap states in the $\pi$ phase.
This correspondence based on the separation of electron and hole-like
excitations is specially clear in the large $\Delta$ limit which we discuss in
what follows.

{\it $\Delta \rightarrow \infty$ case:} As shown in previous works \cite{vecino,hewson,simon} the problem can be 
exactly diagonalized in this limit which already illustrates in a simple way the 0-$\pi$ transition. 
The states can be classified according to the total spin $S=0$ or $S=1/2$. In the $S=1/2$ sector
the energy levels are simply $\epsilon_0$ (doubly degenerate) while in the 
$S=0$ case the states
are given by $E_{0,\pm} = \epsilon_0 +  U/2 \pm \sqrt{(\epsilon_0+U/2)^2 + \Gamma^2 \cos^2{\phi/2}}$,
leading to a phase transition for $\epsilon_0 > E_{0,-}$. Thus, in the 
$\pi$-phase
the spectral density contains four ABS located at 
$\pm U/2 \pm \sqrt{(\epsilon_0+U/2)^2 + \Gamma^2 \cos^2{\phi/2}}$.  

It is quite straightforward to see that this spectrum is recovered exactly by 
the HFA. 
Indeed, in this limit the self-consistent HF solution is
\begin{eqnarray}
\hat{G}^{r,HF}_{\sigma}(\omega) \rightarrow \hspace{6cm} \nonumber\\ 
 \left( \begin{array}{cc} \omega - \epsilon_0 + (2\sigma-1) \frac{U}{2}
 &  \Gamma \cos{\frac{\phi}{2}}  \\
\Gamma \cos{\frac{\phi}{2}} & \omega + \epsilon_0 + (2\sigma+1) \frac{U}{2}  
\end{array} \right)^{-1} 
\end{eqnarray}
which exhibits the same spectrum as the exact solution. 
As in the $\pi$-phase $U/2 > \sqrt{(\epsilon_0+U/2)^2 + \Gamma^2 \cos^2{\phi/2}}$
the excitations for $\sigma=1/2$ have a hole-like character while those for
$\sigma=-1/2$ have an electron character.
Therefore, it is not surprising that this approximation provides a rather good 
description of the ABS spectrum in the $\pi$-phase for the full model.

\section{ABS spectrum for the S-QD case}
\label{S-QD}

Before discussing the general case with two S leads and fixed phase difference it is worth
analyzing the simpler case of an Anderson impurity coupled to a single BCS lead. This model exhibits
also a transition to a degenerate $S=1/2$ ground state and its spectral properties are relevant
to understand the transport properties in N-QD-S systems when $\Gamma_N \ll \Delta$ \cite{deacon}.
To obtain the numerically exact ABS spectrum for this case we have implemented an NRG algorithm
following the lines of Refs. \cite{japs,choi2,karrasch}. 
The idea behind the method is
to discretize the energy levels in the leads on a logarithmic grid of 
energies
$\Lambda^{-n}$ (with the dimensionless parameter
$\Lambda > 1$ and $1\le n \le N \rightarrow \infty$) 
with exponentially
high resolution on the low-energy excitations. 
This discretization allows then to map
the impurity model into a linear ``tight-binding" chain with hopping matrix 
elements
decaying as $\Lambda^{-n/2}$ with increasing site index $n$. 
The sequence of Hamiltonians
which is constructed by adding a new site in the chain is then diagonalized 
iteratively.
As the number of states grows exponentially an adequate truncation scheme 
is required.

The cutoff parameter $\Lambda$ (as defined
originally in Ref. \cite{krishna}) is chosen in order to ensure convergence of 
the spectral properties
inside the gap. Depending on the value of $U/\Gamma$ and the ratio 
$T_K/\Delta$, 
we have chosen $\Lambda$ varying between 2 and 4. For most of the results
shown below we have checked that
the value $\Lambda = 4$ provides already well converged results. 
In all the cases the usual
correction $\Gamma_{NRG} = A_{\Lambda} \Gamma$, where 

\[ A_{\Lambda} = \frac{1}{2} \frac{1 + 1/\Lambda}{1 - 1/\Lambda} \ln \Lambda \]
is used in order to correctly reproduce the exact $\Lambda \rightarrow 1$ limit \cite{krishna,japs,karrasch}.
On the other hand, the maximum number of states $N_c$ kept in the iterative NRG procedure vary
between $\sim 300$ in the S-QD case to $\sim 800$ for the S-QD-S case.

\

\begin{figure}[!t]
\begin{center}
\vspace{0.5cm}
\includegraphics[scale=.35,angle=0]{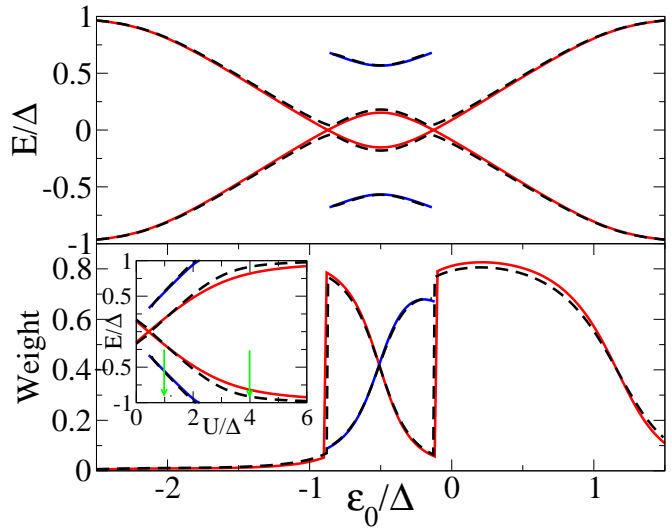}
\caption{(Color online) Upper panel: ABS spectrum of the S-QD system for the 
case $U/\Delta=1$ and $\Gamma/\Delta=0.2$ as a function
of the dot level position $\epsilon_0$, calculated using the NRG 
(full lines) and HFA (dashed lines) respectively. The $\pi$-phase 
appearing around $\epsilon_0/\Delta \sim -0.5$ is characterized by the
crossing of the internal Andreev levels and the presence of two extra 
states around $E \sim \pm 0.5 \Delta$, which disappear within the 0-phase.
The lower
panel depicts the corresponding weights in the spectral density
for the electron-like 
quasiparticle excitations. The inset shows the behavior of the
ABS as a function of $U/\Delta$ in the electron-hole symmetric case $\epsilon_0 = -U/2$. The arrows in this plot indicate the
$U/\Delta$ cases shown in the main frames of Figs. 
\ref{figure1} and \ref{figure2} respectively.} 
\label{figure1}
\end{center}
\end{figure}

In typical experiments the charging energy $U$ adopts a nearly fixed 
value $U>\Gamma$
while the dot level position can be
varied. We thus first analyze the evolution of the ABS spectrum as a function of $\epsilon_0$
for fixed $U/\Delta$ and $\Gamma/\Delta$. Fig. \ref{figure1} shows the ABS spectrum and 
the corresponding weights for $U/\Delta=1$ and $\Gamma/\Delta=0.2$ obtained both within the
NRG method and the HFA. The weights of the spectral density are calculated from the residues of the Green's functions (Eqs. (\ref{lehmann}) and (\ref{hfa})) 
at the poles corresponding to the ABS energies.

\begin{figure}[!t]
\begin{center}
\vspace{0.5cm}
\includegraphics[scale=.35,angle=0]{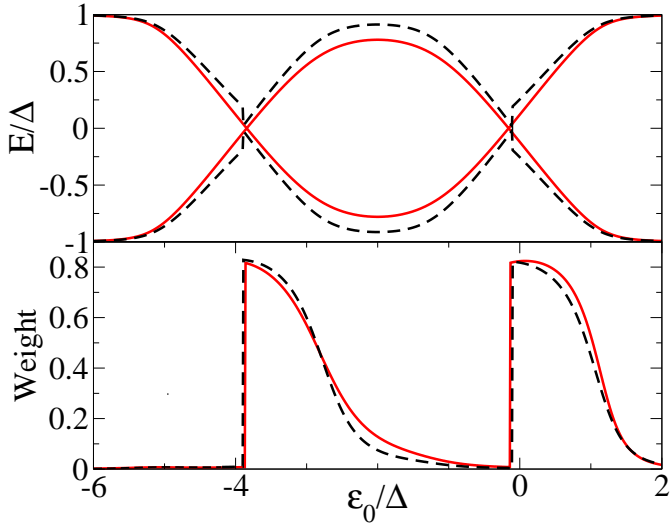}
\caption{(Color online) Upper panel: ABS spectrum of the S-QD
system for $U/\Delta=4$ and $\Gamma/\Delta=0.2$ as a function of the dot
level position $\epsilon_0$ calculated using the NRG (full lines) and
HFA (dashed lines) respectively. Notice that the outermost Andreev bound 
states have already merged with the continuum for this choice of parameters. 
The lower panel depicts the corresponding spectral weight for the electron-like
excitations.} 
\label{figure2}
\end{center}
\end{figure}

It is first worth noticing that the spectrum is characterized by
the presence of 4 ABS in the region $|\epsilon_0+U/2|<U/2$, which corresponds to the $S=1/2$
ground state, while only two ABS are present outside this region where the ground state is 
a singlet. As can be observed the HFA fairly reproduces not only the level positions but also
their weights. Notice that the weights represented in the lower panels of Figs. \ref{figure1}, \ref{figure2}
correspond only to the electron like excitations
which explains the asymmetry between positive and negative $\epsilon_0$ values.

\

\begin{figure}
\begin{center}
\vspace{1cm}
\includegraphics[scale=.35,angle=0]{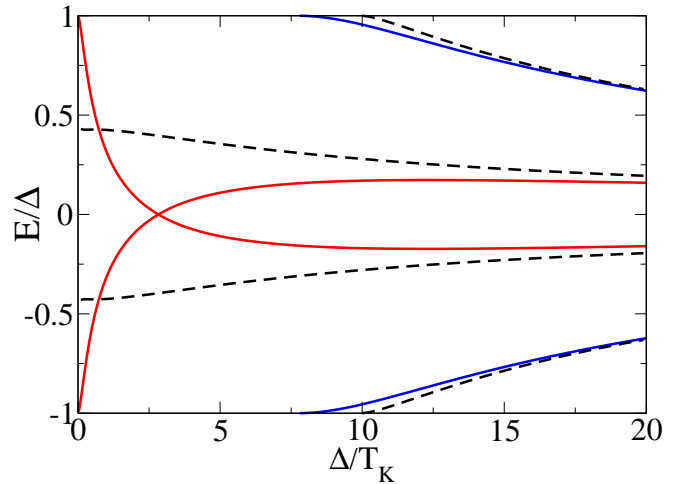}
\caption{(Color online) ABS spectrum of the S-QD system as a function of $\Delta/T_K$, where $T_K$ is the Kondo temperature
for the electron-hole symmetric case taking $U/\Gamma = 5$. Red and blue (full) lines correspond to the NRG calculation
while black (dashed) lines are the HFA results.} 
\label{figure3}
\end{center}
\end{figure}

With increasing $U$ the outermost ABS within the $S=1/2$ phase gradually approach the gap 
edge while its weight is reduced. Eventually, these states disappear for $U/\Delta \sim 2$,
as can be noticed in the inset of Fig. \ref{figure1} which corresponds to the symmetric case.
The ABS spectrum properties for $U/\Delta=4$, illustrated in Fig. \ref{figure2}, clearly
exhibits only two ABS within the gap. Again, as in the $U/\Delta=1$ case the agreement
between the NRG results and the HFA is quite satisfactory. The main difference
between both results appears at the crossing points between the magnetic
and non-magnetic regions where the HFA exhibits a small discontinuity. This
discontinuity is due to the coexistence around these points of both types
of solution in the HFA. In the results represented in Figs. \ref{figure2} and
\ref{figure3} only the most stable HFA solution is shown.

As a general remark one could state that the HFA reproduces fairly well the NRG results
for arbitrary dot occupancy as far as the Kondo temperature $T_K=\sqrt{U\Gamma/2} \exp{(-\pi U/8\Gamma)}$ of the e-h
symmetric case is smaller than $\Delta$. Deviations with respect to the NRG results
could be expected when $\Delta/T_K$ becomes sufficiently small. This is illustrated in Fig. \ref{figure3}
where the ABS spectrum is shown for the e-h symmetric case as a function of $\Delta/T_K$.
As can be observed, when $\Delta/T_K < 5$ the HFA results clearly deviates from the 
NRG ones as it predicts a magnetic solution up to the limit $\Delta/T_K \rightarrow 0$
(i.e. deep in the Kondo regime) whereas the NRG result becomes non-magnetic for $\Delta/T_K \sim 2.8$. 
In contrast, in the opposite limit $\Delta/T_K \gg 1$ both results converge asymptotically to the
$\Delta \rightarrow \infty$ spectrum discussed in the previous section. 
We should point out that for the $U/\Gamma$ ratio used in Fig. \ref{figure3} 
the ``universal" limit (i.e. where all quantities depend only on the ratio 
$\Delta/T_K$)
is still not reached. For larger $U/\Gamma$ ratios the transition occurs 
at smaller 
$\Delta/T_K$, converging to a value $\sim 1.7$ when 
$U/\Gamma > 10$. This value is similar to the one reported in \cite{japs}
but somewhat larger than the one of Ref. \cite{siano} obtained
using quantum Monte Carlo techniques. One should notice also the difference
in the definition of $T_K$ used in Refs. \cite{japs,hewson}, which corresponds
to $0.4107$ times the one used in the present work and also in 
Refs. \cite{choi2,siano,choi,karrasch} (the relation between the two
definitions can be found in \cite{hewson-book}). 

\section{ABS spectrum for the S-QD-S case}
\label{S-QD-S}

\begin{figure}
\begin{center}
\includegraphics[scale=.4,angle=0]{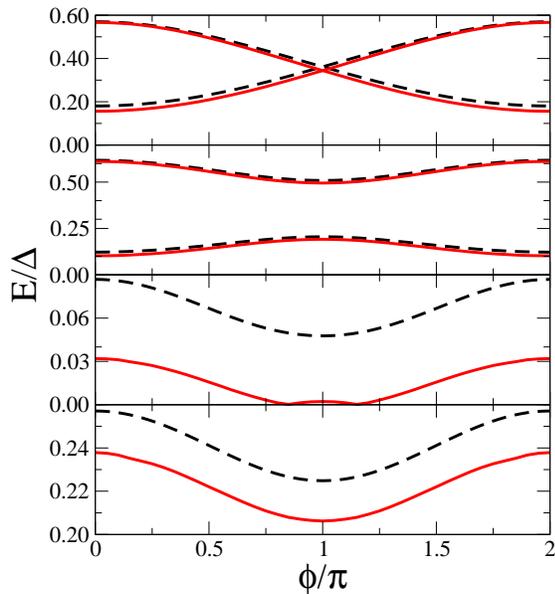}
\caption{Evolution of the phase-dependent ABS spectrum when varying the dot 
level position for $U/\Delta=1$ and $\Gamma/\Delta=0.2$ calculated using
the NRG method (full lines) and the HFA (dashed lines). 
From upper to lower panel $\epsilon_0/\Delta = -0.5, -0.7, -0.95, -1.2$.
Notice that only the electron part of the spectrum is shown and that
a different scale on the vertical axis is used for each panel in order
to better visualize the details of the curves.} 
\label{figure4}
\end{center}
\end{figure}

We analyze in this section the behavior of the ABS spectrum as a function of the phase
difference for the S-QD-S case. The results shown in Fig. \ref{figure4} correspond to
the case $U=\Delta$ and $\Gamma=0.2\Delta$, already analyzed in the previous section, for
different values of the dot level illustrating the transition from the $\pi$ to the $0$
phase. The upper panel of Fig. \ref{figure4} corresponds to the electron-hole symmetric case
where the system exhibits 4 ABS inside the gap (notice that in the figure only the two
electron-like states are shown). The agreement between NRG and HFA is in this regime
fairly good, as was already evident in Fig. \ref{figure1} (which corresponds to the
$\phi=0$ in this plot). When traversing the transition (middle panel of Fig. \ref{figure4})
the agreement is less satisfactory due to the fact that the HFA result fully corresponds
to the 0 phase whereas within NRG the system is in a mixed 0' state (i.e.
a mixed phase of $0$ character at $\phi=0$ and $\pi$ character at $\phi=\pi$
with the absolute minimum energy corresponding to $\phi=0$, see
Ref. \cite{arovas}). Finally, when the level position is sufficiently low
both approaches predict a 0 phase and the agreement in the ABS spectrum 
becomes progressively
quite satisfactory again. This is the same trend which can be observed in Figs. \ref{figure1}
and \ref{figure2}.

\section{Conclusions}
\label{conclusions}

In this work we have analyzed the subgap spectral density of a single dot coupled to 
superconducting leads with the aim of clarifying some of the features of the ABS
which appear to be controversial in the literature. By means of numerically exact NRG
calculations we have shown that in general up to 4 ABS appear when the ground state
becomes magnetic, i.e. in the $\pi$-phase. Within this phase the four states eventually
reduce to only two for increasing $U/\Delta$.
Although the states are located symmetrically
with respect to the Fermi level the electron-hole symmetry is in general broken. We 
have shown that this behavior is adequately reproduced by 
the HFA for a broad range of
parameters, except very close to the transition regions between the different 
phases.
This approximation, however, is unable to reproduce the correct behavior in the
strong Kondo regime when $T_K \gg \Delta$.

\begin{acknowledgements}      
We thank M. Goffman and P. Joyez for their comments on the manuscript.
Financial support from Spanish MICINN through
project FIS2008-04209 and FP7 project SE2ND is acknowledged.
\end{acknowledgements}

\end{document}